 \documentclass[preprint2]{aastex}

\def\ie{{i.e.~}}

\def\eV{e\kern-.15em V}                 % define eV macro
\def\keV{ke\kern-.15em V}                % define keV macro

\def\rej{RE~J1034+396}

\def\lae{\mathrel{<\kern-1.0em\lower0.9ex\hbox{$\sim$}}}
\def\gae{\mathrel{>\kern-1.0em\lower0.9ex\hbox{$\sim$}}}

\def\euve{{\sl EUVE}}
\def\sax{{\sl Beppo-SAX}}
\def\rosat{{\sl ROSAT}}
\def\asca{{\sl ASCA}}

\def\hst{{\sl HST}}

\shorttitle{Puchnarewicz et al.}
\shortauthors{Black hole mass and accretion rate in \rej}

\begin{document}


\title{Constraining the black hole mass and accretion rate in the narrow-line 
Seyfert~1 \rej}


\author{E. M. Puchnarewicz and K. O. Mason}
\affil{Mullard Space Science Laboratory, Holmbury St. Mary, Dorking,
Surrey RH5 6NT, UK}
\email{emp@mssl.ucl.ac.uk, kom@mssl.ucl.ac.uk}

\author{A. Siemiginowska and A. Fruscione}
\affil{Harvard-Smithsonian Center for Astrophysics, 60 Garden
Street, MS-4, Cambridge MA 02138, USA}
\email{aneta@head-cfa.harvard.edu, antonell@head-cfa.harvard.edu}

\author{A. Comastri}
\affil{Osservatorio Astronomico di Bologna, via Ranzani 1, 
I-40127 Bologna, Italy}
\email{comastri@astbo3.bo.astro.it}

\author{F. Fiore\altaffilmark{1}}
\affil{Osservatorio Astronomico di Roma, via Frascati 33, 
I-00040 Monteporzio (Rm), Italy}
\email{fiore@quasar.mporzio.astro.it}

\and

\author{I. Cagnoni\altaffilmark{1}}
\affil{Scuola Internazionale Superiore di Studi Avanzati, 
via Beirut 2-4, 34014 Trieste, Italy}
\email{ilaria@head-cfa.harvard.edu}


\altaffiltext{1}{Harvard-Smithsonian Center for Astrophysics, 60 Garden
Street, MS-4, Cambridge MA 02138, USA}


\begin{abstract}

We present a comprehensive study of the spectrum of the narrow-line
Seyfert 1 galaxy \rej, summarizing the information obtained from the
optical to X-rays with observations from the William Herschel 4.2m
Telescope (WHT), the Hubble Space Telescope (\hst), the Extreme
UltraViolet Explorer (\euve), \rosat, \asca\ and \sax. The
\sax\ spectra reveal a soft component which is well-represented by two
blackbodies with $kT_{\rm eff}$=60~\eV\ and 160~\eV, mimicking that
expected from a hot, optically-thick accretion disc around a low-mass
black hole. This is borne out by our modeling of the optical to X-ray
nuclear continuum, which constrains the physical parameters of a NLS1
for the first time. The models demonstrate that \rej\ is likely to be
a system with a nearly edge-on accretion disk (60 to 75$^\circ$ from
the disk axis), accreting at nearly Eddington rates (0.3 to 0.7
$L_{\rm Edd}$) onto a low mass black hole (M$_{bh}\sim$2 to 10$\times 10^6$
M$_{\odot}$). This is consistent with the hypothesis that NLS1s are
Seyfert-scale analogies of Galactic Black Hole Candidates. 
The unusually high temperature of the
big blue bump reveals a flat power-law like continuum in the
optical/UV which is consistent with an extrapolation to the hard X-ray
power-law, and which we speculate may be similar to the continuum
component observed in BL Lac objects in their quiescent periods. From
the \sax\ and \asca\ data, we find that the slope of the hard X-ray
power-law depends very much on the form of the soft component which is
assumed. For our best-fitting models, it lies somewhere between
$\alpha$=0.7 and 1.3 and thus may not be significantly softer than AGN
in general.

\end{abstract}


\keywords {accretion, accretion disks -- galaxies: active -- galaxies:
nuclei -- galaxies: Seyfert -- X-rays: galaxies -- 
galaxies: individual (\rej)}

\section{Introduction}

The narrow-line Seyfert 1 (NLS1) type of AGN has proven to be a
valuable and a fascinating resource for the study of the class as a
whole. They have very strong soft X-ray excesses \citep{puc92,bol96},
are often highly variable \citep{bra96,bol97} and, in some
cases, the optical/UV big blue bump (BBB) component is so hot that it
is shifted into the UV/EUV regime, leaving a bare, power-law-like
continuum component in the optical \citep{puc95}.

One of the earliest hypotheses put forward to explain the strong 
and variable ultra-soft X-ray excesses, was a high
mass accretion rate onto a relatively low-mass black hole
\citep{pou95}. 
This was originally proposed for \rej, by analogy with
the properties of Galactic Black Hole Candidates (GBHCs). An
alternative model was that such systems were geometrically-thick
accretion disks (ADs) viewed face-on, lying co-planar with a flattened
broad line region [BLR; \citet{puc92}].  

In our initial study of \rej\ \citep{puc95}, we showed
that the overall IR to X-ray spectrum compared well with the
combination of a geometrically-thin, optically-thick AD and an
underlying power-law with a slope, $\alpha$=1.3 ($\alpha$ is defined
throughout such that $F_\nu\propto\nu^{-\alpha}$). The Kerr black hole
mass was 7$\times10^5$ M$_\odot$, the accretion rate was 0.073
M$_\odot$ yr$^{-1}$ and the disk was viewed at an angle of 60$^\circ$
from its axis. However, this was for illustrative purposes only and no
constraints were placed on these parameters. Thus, while this might
have suggested the presence of a low-mass black hole for \rej,
consistent with the GBHC analogy, solutions at higher black hole mass
could not be ruled out.

With the launch of {\sl Beppo-SAX} and the inclusion of \rej\ in the
NLS1 core program, came the opportunity to place meaningful
constraints on black hole mass (M$_{bh}$), accretion rate (\.M) and
inclination. \rej\ was also due to be observed with {\sl HST} within a few
months of {\sl Beppo-SAX}, and an optical spectrum was also scheduled
in {\sl WHT} service time within a few weeks. Thus a
quasi-simultaneous spectrum, providing the most complete coverage
possible from the optical to mid X-rays was available. Simultaneity is
important for a NLS1 due to the variable nature of the class,
although \rej, contrary to its counterparts, has shown remarkable
stability. Thus these data have provided the best opportunity yet to
fit AD models and provide constraints on the defining parameters of an
NLS1. Using a combination of a power-law and an optically-thick,
geometrically-thin Kerr AD model, we present fits to the observed
spectrum of \rej, measuring M$_{bh}$ and \.M and constraining the
inclination for the first time, and discuss the results in the context
of the GBHC model.

The improved UV and optical data presented here have also allowed us
to re-visit the issue of what is producing the optical/UV continuum in
this object. Our previous work had suggested a power-law-like
continuum rising towards the red, which was not due to the host galaxy
or the BBB. We have separated out the galaxy component from the
spatially-resolved optical spectrum obtained and, when combined with
the {\sl HST} data, examined the pure nuclear component with a greater
degree of clarity. The results and their implications for the
production of the optical continuum in AGN are also discussed in this
paper.

\section{Data}

\rej\ has been observed on several occasions at different wavelengths from the
radio to X-rays \citep{puc95,pou95,puc98,bre98}. In this paper, we
present data taken at optical, EUV ({\sl EUVE}), soft X-ray ({\sl ROSAT} HRI
and {\sl SAX} LECS) and medium X-ray ({\sl SAX} LECS and MECS) energies which
were taken  quasi-simultaneously with the {\sl HST}-FOS spectrum. We also
analyze previously unreported {\sl ROSAT} HRI (taken 1994 November 20) and {\sl
ASCA} data (taken 1995 May 18) to check for variability in this object.

\subsection{Optical spectra}

\rej\ was observed on the night of 1996 March 24 using the twin-armed 
ISIS spectrograph on the William Herschel Telescope on La Palma as part of 
the service observing program (observers D. Pollacco and D. King). The 
input starlight was divided between the red and blue arms of the 
spectrograph by means of a dichroic filter whose crossover wavelength is 
5400\AA. The R316R grating was used in the red arm of the spectrograph, 
centered on a wavelength of 6677 \AA. The blue arm utilized the R300B 
grating centered on 4519 \AA. The red spectrum spanned the wavelength range 
6000\AA\ to 7700\AA, while the blue spectrum extended from 3800\AA\ to 
5300\AA. The data were recorded on TEK 1024 square CCDs. Two exposures, 
each of 600s duration, were taken of \rej\ and the resulting images 
combined during analysis (eliminating cosmic ray events) to give a total 
effective exposure time of 1200s. The sky was clear throughout. Standard 
exposures of wavelength calibration arcs and flux calibration stars were 
made immediately following the \rej\ observations using the same 
instrumental set-up.

The integrated spectrum of \rej\ is clearly contaminated by light from 
the host galaxy, particularly in the red. The separate contributions of the 
active nucleus and the underlying galaxy are clearly evident in the 
cross-dispersion spatial profile. To remove the worst effects of the galaxy 
contamination on the nuclear spectrum, the 2-D image of the spectrum was 
divided into 25~\AA\ bins and the mean spatial profile in the 
cross-dispersion direction formed for each bin. Two Gaussian profiles 
superimposed on a flat background were then fitted simultaneously to the 
spatial data in each bin, to model the nuclear and extended components 
respectively. Gaussian functions were found to be satisfactory and robust 
representations of the spatial data for the purposes of this decomposition. 
The flux in the two Gaussian components as a function of wavelength thus 
represents the spectrum of the nuclear and extended emission respectively.

\subsection{{\sl HST} observations}

\rej\ was observed by {\sl HST} on 1997 January 31 using three gratings (G130L,
G190L and G270L) covering the range 1100~\AA\ to 3300~\AA. Full details of the
observations and data reduction were presented in \citet{puc98}.

The {\sl HST} spectrum was combined with the de-convolved optical
nuclear spectrum and the resulting optical/UV continuum was fitted
with a power-law model. The best fit was obtained for $\alpha$=0.9,
having removed obvious emission and absorption features before
fitting.  Errors are difficult to determine precisely, because the
relative normalization of the two spectra is not known (the data were
not taken simultaneously and different slit widths and positions were
used) and because of the difficulty in securely removing absorption
and emission features before fitting the continuum. Assuming,
therefore, a conservative error of 20\% on the optical and
UV fluxes, gave a greater than 99.9\% probability that the
model was a good fit to the data.

\subsection{{\sl EUVE} data}

\rej\ was observed with the Deep Survey Spectrometer (DS/S) on board \euve\ 
from 14 April 1997 UT  05:35:04 to 20 April 1997 UT 01:03:07
for a total of approximately 100 ks. The observation was part of a 
simultaneous \euve/\sax\ campaign.

The DS/S \citep{wel90} is equipped with a broad band imaging
detector (covering the $66-178$ \AA\, or $0.07-0.18$ keV band in the
Lexan/B filter) and three spectrometers \citep{het83,abb97} covering
the ``short'' (SW: $70-190$ \AA, $0.06-0.18$ \keV), ``medium'' (MW:
$140-380$ \AA) and ``long'' (LW: $280-760$\AA) EUV wavelengths.  This
configuration allows simultaneous imaging and spectroscopy with a
spatial resolution of $\sim 1$ arc min and a spectral resolution of
$\lambda/\Delta \lambda \sim 200$ at the short wavelengths.

\rej\ was observed $0.2^\circ$ off-axis in order to avoid the DS
dead-spot
and to extend the spectrum toward shorter wavelengths
($\lambda_{\rm{min}}\sim 66$ \AA) with respect to a regular on-axis
observations ($\lambda_{\rm{min}}\sim72$ \AA).
[The DS dead-spot is a small region of reduced
gain and detector quantum efficiency near the center of the DS
detector, caused by the observation of the very bright EUV source
HZ~43 \citep{sir93}.] 

\subsubsection{Light-curve}

We extracted the light-curves from the DS time-ordered event list
using the \euve\ Guest Observer Center software (IRAF/EUV package) and
other IRAF timing tasks adapted for \euve\ data.  After correcting the
data for instrumental dead-time, telemetry saturation, vignetting and
eliminating intervals of high particle background, the total effective
exposure is 97061~s.

We counted the source photons in a circle of $1.1^\prime$
radius, and we estimated the background in a concentric annulus with
inner and outer radii of $1.5^\prime $ and $3.8^\prime$
respectively. The extraction region includes more than 98\% of the DS
point spread function \citep{sir97}.  The effective area of the
DS instrument (25~cm$^2$ at $\lambda=85$~\AA), is more than 10 times
larger than the spectrometer effective area at this wavelength and a
good detection of the source (average signal-to-noise ratio (SNR)
$\sim 4.9$) was obtained during each orbit.  Figure 2 shows the light
curve binned over one average \euve\ orbit (about 5544~s). Since the
\euve\ instruments are shut down during satellite daytime (around 2/3
of the total orbital time for EUVE 550~km circular orbit with an
inclination of $28^\circ$) and passages through regions of high
particle background (such as the South Atlantic Anomaly) the average
effective exposure per orbit is approximately 1000~s, after discarding
the bins with effective exposure less than 200~s.

The average count rate over the entire observations is $0.038 \pm
0.001$~counts s$^{-1}$.  As a comparison \rej\ was observed twice
before by \euve: during the all-sky survey in 1992-1993 it had a count
rate of $0.023\pm 0.007$ in $\sim 1600$~s of observation
\citep{fru96} and again serendipitously on 27 December 1993 for an
effective exposure time of $\sim 5000$~s; the count rate then was
$0.029 \pm 0.003$ (Fruscione, private communication). During the
observation described in this paper (April 1997), \rej\ was in a
relatively bright state, $\sim$60\% brighter than in the
all-sky survey and $\sim$30\% brighter than in December 1993.

As immediately visible from the light-curve, the EUV flux from \rej\
remained effectively constant for the 5 days of observation and in fact a
$\chi^2$ test for a constant source on the entire data set (excluding
intervals with effective exposure below 200~s) gives a value of 51 for
75 degrees of freedom, corresponding to a $\sim 1$\% probability that
the source varied during the six days of this observation.

\subsection{{\sl ROSAT} HRI images}

Since the re-discovery of the active nucleus in \rej\ by the {\sl ROSAT} Wide
Field Camera in 1990, several observations have been made in the soft and
medium X-rays and a log of the available data to date is given in Table 1.

A measurement of the soft X-ray flux from \rej\ was
made with the HRI on {\sl ROSAT} in 1996 November and yielded a count
rate of 0.61$\pm$0.02 count s$^{-1}$ with no evidence for variability
for the duration of the observation ($\sim$30~ks). The full width at
half maximum (FWHM) of the profile of \rej\ was 7 arc sec. The point
spread function (PSF) of an unresolved on-axis source observed by the
HRI has a FWHM between $\sim$5.1 and 7.4 arc sec, thus the soft X-ray
emission of \rej\ in the {\sl ROSAT} HRI was consistent with a point source.

A previous HRI observation of \rej\ was made in 1994 November 20. We
measure a count rate of 0.49$\pm$0.02 from these data, thus the soft
X-ray flux had increased by $\sim$20\% in the two years since
the 1994 pointing. This is consistent with {\sl EUVE} observations which
showed an increase between 1992/1993 and 1997 (see
Section 2.3.1).

\subsection{{\sl ASCA} observations}

\rej\ has been observed on two occasions by {\sl ASCA}, in 1994
November (coincident with the first {\sl ROSAT} HRI observation) and
six months later in 1995 May. The first dataset was analyzed by
\citet{pou95} but the second is unpublished to date. We present the
1995 data here and re-analyze the 1994 spectra to ensure a consistent
approach towards the data reduction when comparing observations.

Both sets of SIS observations were taken in 1-CCD mode. The GIS
source spectra were extracted using a circle of radius 4 arc min. The
GIS background regions were selected from a position opposite that of
\rej, \ie at the same radial distance, and with a circle of radius 6.7
arc min. The SIS spectra were extracted using a circle of radius 3
arc min and a nearby circular region of the SIS chips was used for
the background. The background light-curves were used to check for and
discard any periods of anomalous data before the spectra were extracted. 

The background subtracted light-curves of the target were examined for
signs of significant variability. None was seen in the 72020000 data
[consistent with the results of \citet{pou95}]. In the 72020010
observation, there is a suggestion of variability at the 20-30\%
level, although a $\chi^2$ test indicates that this cannot be
distinguished from flux at a constant level. The count rate of \rej\
in 1995 May was consistent with that six months earlier in 1994
November at the 2$\sigma$ level.

The reduced spectra were fitted with a combination of power-law and
thermal models, using the {\sc xspec} spectral fitting package
(Version 11.00) and the results are given in Table 1. The fitting
ranges were 1.0 to 10~\keV\ for the GIS and 0.7 to 10~\keV\ for the
two SIS. In all the fits, a Galactic column density of 1.5$\times
10^{20}$ cm$^{-2}$, interpolated from the maps of \citet{sta92}, was
assumed. For the earlier 72020000 observation, the lowest
$\chi_\nu^2$s were obtained for the two power-law and two blackbody
plus power-law fits (full details are given in the table). For all
fits, the best fit intrinsic column density converged to 0, indicating
little if any absorption local to the AGN. The slopes of the hard
X-ray power-law for the two best-fitting models were
$\alpha\sim$0.7-0.8 which is typical of AGN in general
\citep{mus84,com92} and {\sl not} significantly softer as suggested
by \citet{pou95}.

The second dataset, 72020010, was observed for only one-third as long
as the earlier one, thus the fits are more poorly constrained. Only a
single power-law model gave an acceptable fit to the data if the
intrinsic column was allowed to be free: for the other three models
tried, the intrinsic column converged to a few times $10^{21}$
cm$^{-2}$. Such high column densities were in turn offset by
un-physically strong un-absorbed components, suggesting some flattening
of the spectrum towards lower energies. A broken power-law model with
no intrinsic column did give an improved fit over a single
power-law. The single power-law fit could not be improved by the
addition of a bremsstrahlung or blackbody component however.

The 72020010 data were well-fitted with the best-fitting two blackbody plus
power-law model for the earlier data, giving a $\chi_\nu^2$=0.95 for
156 degrees of freedom (ie having fixed all parameters except the
normalizations). This suggests that there was no significant spectral
variability either between the two {\sl ASCA} observations.

\subsection{{\sl Beppo-SAX} observations}

\rej\ was observed on 1997 April 18 and 19 with the \sax\ Narrow
Field Instruments, the Low Energy Concentrator Spectrometer [LECS;
0.1-10~\keV\ \citep{par97}], the Medium Energy Concentrator
Spectrometer [MECS; 1.3-10~\keV\ \citep{boe97}], the High
Pressure Gas Scintillation Proportional Counter [HPGSPC; 4-60~\keV\
\citep{man97}] and the Phoswich Detector System [PDS; 13-200~\keV\
\citep{fro97}]. The source was not detected in the high energy
instruments thus we report here the analysis of the LECS and MECS
only.

The observations were performed with all three MECS units and data
from these were combined together after gain equalization.  The LECS
is operated during dark time only, therefore LECS exposure times are
usually smaller than MECS ones by a factor 1.5-3 (a factor of 2 for
\rej).  Standard data reduction was performed using the SAXDAS
software package (\url{www.sdc.asi.it/software/saxdas}).  In
particular, data are linearized and cleaned from Earth occultation
periods and unwanted periods of high particle background (satellite
passages through the South Atlantic Anomaly). The LECS, MECS and PDS
backgrounds are relatively small and stable (variations of at most
30\% during the orbit) due to the satellite's low inclination orbit
(3.95$^\circ$). Therefore data quality depends little on screening
criteria such as Earth elevation angle, Bright Earth angle and
magnetic cut-off rigidity (we accumulated data for Earth elevation
angles $>5^\circ$  and magnetic cut-off rigidity $>6^\circ$).

We extracted the LECS spectrum from 8 arc min radius regions.  This
radius maximizes the signal-to-noise ratio below 1 keV in the
LECS. The MECS image is {\sl not} consistent with the instrument PSF
and the spatial analysis will be presented elsewhere (Fiore et
al., in preparation). We have used source extraction radii of 2
and 3 arc min to avoid as much as possible problems of confusion or the
contribution of extended emission. The results for 2 and 3 arc min
extraction radii were consistent with each other and we report in the
following, those obtained with the 3 arc min extraction radius.

The LECS and MECS internal backgrounds depend on the position [see
\citet{chi98}, and the BeppoSAX Cookbook,\\
\url{www.sdc.asi.it/software/cookbook}].  Background spectra have then
been extracted from high Galactic latitude `blank' fields from regions
equal to the source extraction region, in detector coordinates. We
have checked whether the mean level of the background in these
observations is comparable with the mean level of the background in
the quasar observations using source free regions at various positions
in the detectors.  In the LECS the mean local background is consistent
with the `blank' field mean background.  In the MECS we found that the
local background is higher than the blank field background by 7\%. We
have verified that the excess is constant in energy.  We then scaled
by the same amount the blank field background spectra before
subtraction.

Table 3 gives the observation dates, LECS and MECS exposure times and
the detected count rate in each instrument.

Spectral fits were performed using the XSPEC 9.0 software package and
public response matrices as from the 1997 August 31$^{st}$ issue.
Spectra were always re-binned following two criteria: a) to sample the
energy resolution of the detectors with four channels at all energies
where possible, and b) to obtain at least 20 counts per energy
channel. This allows the use of the $\chi^2$ statistic in determining
the best fit parameters, since the distribution in each channel can be
considered Gaussian while avoiding possible systematic errors due to
the linearity in the spacing of the original sampling. Constant
factors have been introduced in the fitting models in order to take
into account the inter-calibration systematics between instruments
\citep{fio98}.  The expected factor between LECS and MECS is about
0.9.  In the fits we then assume the MECS as reference instruments and
constrained the LECS parameter to vary in the small range 0.7-1.1.
The energy range used for the fits are: 0.1-4 keV for the LECS
(channels 11-400), 1.65-10 keV for the MECS (channels 37-220)

\subsubsection {Spectral analysis}

LECS and MECS spectra were fitted with a number of models: single
power-law, broken power-law, 2 power-laws and power-law plus disc black
body. In all cases the emission spectrum was reduced at low energy by
a cut-off due to neutral gas along the line of sight. The column
density of the gas was always constrained to be greater than or equal
to the Galactic value along the line of sight of $1.5\times10^{20}$
cm$^{-2}$.

Table 4 gives the best fit parameters along with the $\chi^2$
obtained in each case.  To ease the comparison of the {\sl Beppo-SAX}
observation with previous ROSAT and ASCA observations we have also
fitted the single LECS and MECS spectra with a power-law model limited
to the ranges 0.1-2 keV and 2-10 keV.

The 90\% limit on a 6.4~\keV (6.7~\keV) iron K$\alpha$ line in
the {\sl BeppoSAX} data is 165~\eV\ (210~\eV).

\subsubsection{Spectral curvature}
 
Figure 3 shows the fit with a single power-law: positive residuals are
evident above 3-4 keV .  The spectrum is clearly flattening above this
energy. The significance of the flattening is demonstrated in Figure 4
which shows the contour plot for the soft and hard slopes of the
broken power-law model. The figure shows that the soft index,
$\Gamma_{low}$ is softer than the hard, $\Gamma_{high}$ at greater
than 99\% significance.

The fits with two power-laws, a broken power-law and a blackbody plus
a power-law all give a $\chi^2$ which is significantly smaller than a
single power-law (Table 4). The blackbody plus power-law model
produces an acceptable $\chi^2$ and a best fit absorbing column
density lower than Galactic, unlike the other two
models. Therefore either we are measuring a small column in the host
galaxy of \rej, or the emission spectrum has a curvature also at low
energy. While the former case is certainly possible, the possibility
of a low energy turn down of the emission spectrum is intriguing,
because it may suggest a small black hole mass, if the low energy
emission is interpreted in terms of optically thick disk emission.

Indeed the soft X--ray spectrum is best fitted with two blackbodies
which mimic a multi-temperature disc model. This result is consistent
with the analysis of the {\sl ROSAT} PSPC data by \citet{puc95}, who
found a significant hardening of the 0.1-2~\keV\ spectrum at low
energies, when the column density was fixed at the Galactic value. 

The spectrum was also fitted with a bremsstrahlung plus power-law
model which gave a $\chi^2$ of 75.5 (for 94 degrees of freedom). Thus
although a bremsstrahlung form for the soft X-ray component cannot be
ruled out, the two-blackbody representation does produce the lowest
$\chi_\nu^2$. Furthermore, a bremsstrahlung form for the soft X-ray
component is not generally popular, because the implied extent of the
optically-thin gas would be much larger than suggested by the
variability of most NLS1s.

The preferred slope of the hard X-ray power-law continuum
depends on the form assumed for the ultra-soft component. For the
best-fitting two blackbody plus power-law model, the hard power-law is
best represented by a slope $\alpha$=1.2. This is slightly harder than
that measured by \citet{pou95} but only slightly softer than
that of AGN in general (where $\alpha\sim$1). Our analysis of the
\asca\ data suggest a slope for the hard X-ray power-law, $\alpha$,
between 0.7 and 1.3, depending on the model assumed for the soft component. 

Comparing the \sax\ spectrum with the \asca\ data, we find that the
flux and spectral form of all three observations are consistent within
90\%.

\section{Model fits to the optical to X-ray continuum}

In spite of its historical significance as one of the first NLS1s with
a strong ultra-soft component to be reported, \rej, unlike many others
in its class, has exhibited a remarkable degree of flux and spectral
stability. In this paper, we present a comprehensive review of
measurements of \rej\ in the EUV to X-rays over 6 years, and over this
period it has shown changes on long time-scales of up to 60\% (compared
to eg. RE~J1227+164 whose flux changed by a factor of 70). On short
time-scales, no significant changes have been observed. Thus, in a
study such as this where we wish to fit non-simultaneous optical to
X-ray data of a NLS1, \rej\ is the prime candidate. Obviously
simultaneous optical to X-ray data, such as that which will be
provided by the Optical Monitor and EPIC on XMM Newton, would be the
ideal, but until such data are obtained, the quasi-simultaneous
observations presented here provide the best available.

The `nuclear' spectrum of \rej\ is illustrated in Figure 5. This is
made up of the point-source component extracted from the optical data
(Section 2.1), the {\sl HST} spectrum (Section 2.2) and the {\sl
Beppo-SAX} spectrum (Section 2.6).

We have assumed that the optical-UV-X-ray emission originates
primarily in a geometrically thin accretion disk around a rotating
super-massive black hole. The radial temperature of the standard,
Keplerian disk is calculated given the black hole mass and the
accretion rate.  We approximate the disk emission by local blackbody
emission, which is modified by electron scattering in the disk
atmosphere \citep{cze87}. Relativistic corrections have been applied
to the disk emission following \citet{lao89}.

We fit the optical to X-ray spectrum of \rej\\ using a grid of
accretion disk models parameterized by a central black hole mass (M$_{bh}$),
accretion rate (\.M) and inclination angle \citep{sie95}.
Because there is no signature of the BBB in the UV spectrum of
RE~1034+396 we also add a power-law component to the disk emission in
order to describe the optical/UV continuum. We normalized a power
law at 2~\keV\ to 8$\times10^{42}$ erg s$^{-1}$ with a slope fixed
to $\alpha = 1.15 $ to connect the near IR and hard X-ray points (see
Figure 5). The implications of a possible physical significance of this
component are discussed in Section 4.2.

The fact that the spectral turn-over is detected in the soft X-ray
band (Section 2.6.2) can be used to constrain the minimum black hole
mass required for the source.  Assuming that the peak seen in the soft
X-ray spectrum is related to the maximum effective temperature of the
optically thick disk, then the physical radius at which this
emission is generated is given by \citep{fra92}:
\begin{equation}
{\sigma T_{\rm eff}^4} =  {{3GM \dot M} \over {8 \pi R_{max}^3}}
(1- \sqrt {{3R_{Sch}} \over {R_{max}}})
\end{equation}
\begin{equation}
R_{max} = {49 \over 36} R_{Sch}
\end{equation}
where $T_{\rm eff}$ is the effective temperature, $R_{Sch} = 2 G M_{bh}
/c^2$ is the Schwarzschild radius of the black hole 
and $R_{max}$ is the radius where the maximum
temperature is reached.

At a given effective temperature the disk contributes the most of the
emission to the frequency given by \citet{sie89} and \citet{cze87}:
\begin{equation}
kT_{\rm eff} = 1.625 h\nu
\end{equation}
This allow us to link the turn-over observed in the spectrum with the
central black hole mass.  Our approximation assumes that the disk is
optically thick and the entire spectrum is described by local
blackbody emission.  However, at high temperatures the disk
photons are re-scattered by electrons in the atmosphere of the
disk. Some photons gain energy and contribute to the soft X-ray
spectrum. This modification allows to see the turn-over at higher
frequencies, so a higher black hole mass is allowed. 

For a turn-over at $\sim 0.25$keV and an accretion rate
corresponding to the Eddington luminosity, the mass derived from the
above equations is $\sim 10^4$M$_{\odot}$. This mass can be larger if
most of the observed photons are Comptonized in the disk
atmosphere. The UV photons can gain energy up to a mean value of
3$kT$. This is to  $\sim 0.25$~\keV\ for photons at the spectral
peak of \rej. If we take this temperature into account then
the central black hole mass is estimated to be of the order $6 \times
10^6$M$_{\odot}$, again assuming the critical accretion rate.

Using the optical to X-ray data presented here, best-fitting
parameters for the black hole mass and accretion rate have been
derived by constructing $\chi_\nu^2$ grids for models compared with
the data over a range of inclination angles.  In Figure 6 we show the
confidence intervals for accretion rate and black hole mass at
inclinations [cos(inc)] of 0.25, 0.5, 0.75 and 1.0 [where cos(inc)=1.0
corresponds to a face-on disk].  The face-on configuration is
ruled-out: note also that, in general, the highest accretion rates are
not consistent assumptions built into the model. The results suggest
that \rej\ contains a relatively small black hole mass, with
the best fitting value of $\sim 5\times 10^6$M$_{\odot}$.  The
accretion rates are relatively high, but consistent with the model
assumption. The best accretion rate value is found to be equal to
0.4~\.M$_{\rm Edd}$ for a relatively edge-on inclination angle of
cos(inc)=0.25.  In Figures 7 and 8, we show some of the fits to the
nuclear spectrum at cos(inc)=0.25 and 0.5, around the best-fitting
values.

One problem with modeling these data, is that the signal to noise
ratios in the optical and UV, and the number of available data points,
are much higher than that in X-rays. This results in an imbalance in
the significance of the optical/UV data over the X-rays in the
$\chi_\nu^2$ statistic, despite the physical significance of the form
of the X-ray spectrum.

To establish which parameters the high energy data alone would prefer,
we have repeated this fitting procedure using only the X-ray data. The
resulting $\chi_\nu^2$ grids are consistent with those for the full
optical/UV/X-ray data, with a minimum at M$_{bh}=5\times10^6$M$_{\odot}$,
\.M=0.4~\.M$_{\rm Edd}$ and cos(inc)=0.25, and the face-on geometry
ruled out.

As a further check on the robustness of the results, we have also
`matched' the resolution of the optical/UV data to the X-rays. This
was done by binning the spectra so that the number of optical/UV data
points per frequency decade was similar to that in the X-ray
spectrum. Although this reduces the error on the measurement of the
flux in each bin, the overall error remains dominated by uncertainties
in the absolute flux determination and by the effects of any spectral
variability. Thus, the errors on the binned optical/UV data were also
assumed to be $\pm$20\%. Again, the grids for the binned optical/UV
data were consistent with those for the full optical/UV/X-ray data as
above. The $\chi_\nu^2$'s were generally higher at each point in the
grids, but the minima had values of $\sim$1.

\section{Discussion}

The {\sl HST}-FOS data and quasi-simultaneous 0.1-10~\keV\ spectrum
measured by {\sl Beppo-SAX} provide very tight constraints on the form of
what we assume to be the BBB (\ie a single component which would
normally encompass the optical/UV BBB and the soft X-ray excess) in
\rej. No sign of significant emission from the BBB is seen in the FOS
spectrum down to $\sim$1250~\AA, while the LECS spectrum measures a
very steep and relatively strong soft X-ray component. There is also
evidence for a flattening towards the lowest energies in the X-ray
spectrum (below $\sim$0.3~\keV), confirming earlier {\sl ROSAT} PSPC
observations. Thus the unusually high temperature of the BBB in \rej\
makes the source unique for two reasons: {\sl (1)} the BBB is shifted
out of the optical/UV range completely; and {\sl (2)} assuming {\sl
no} cold gas absorption intrinsic to \rej, the high energy turnover of
the BBB is observed in soft X-rays at $\sim$0.25~\ keV.

\rej\ is important in one further respect: it shows no evidence for
significant variability on short time-scales ($\sim$days), and only
moderate variability on long time-scales (months to years). This is
contrary to the nature of Seyferts in general, where sources with very
soft X-ray spectra tend to show large amplitude variability on short
time-scales, while harder sources vary less [see eg. \citet{fio98} and
references therein]. 

\subsection{Black hole mass and accretion rate}

This unique set of observational properties, and the availability of
quasi-simultaneous data, has enabled us to place meaningful
constraints on the essential parameters of \rej, ie. the black hole
mass, accretion rate and inclination of the AD to our line of
sight. The $\chi_\nu^2$ grids favor almost Eddington accretion rates
($\sim$0.4\.M$_{\rm Edd}$) onto a relatively low mass black hole
(M$_{bh}\sim5\times10^6 $M$_\odot$), for a disk inclined relatively edge-on
(ie. 75$^\circ$ from the axis of the disk). This is consistent with
the hypothesis of \citet{pou95}, who suggested that \rej, and
NLS1s in general, were systems analogous to Galactic Black Hole
Candidates (GBHCs). Models of GBHCs predict that, in their
`high-state', the accretion rate onto the black hole is high,
resulting in a strong soft thermal component and a relatively soft,
fluctuating hard power-law. 

If, in the case of NLS1s, the matter was accreting at high rates onto
a relatively low mass black hole, this would also result in a
relatively high temperature for the inner edge of the AD. Thus a very
hot, soft X-ray component would be observed in NLS1s when compared to
other AGN. In addition, the low-mass black hole might result in a
low-velocity broad line region (BLR), either because the clouds were
moving more slowly due to the weaker gravitational potential, or
because the emitting regions were located further out (which in turn
might be due to the weaker gravitational field and/or the unusual form
of the incident ionizing continuum).

Our results are consistent with this idea of high accretion onto a
relatively low mass black hole. There is one possible problem that
should be addressed however, and that is the inclination of the
AD. The fits prefer an almost edge-on disk ($\sim60-70^\circ$), which
is not unexpected as such, since the probability of observing an AD in
soft X-rays, assuming isotropic emission of the soft X-ray flux, is
proportional to the sine of the inclination angle. This in itself is
not a problem, but one might naively expect any molecular torus to lie
roughly co-planar with the AD. Given that the opening angle of such
tori is thought to be about 45$^\circ$ and the optical depths implied
are high [$\sim10^{24}$ cm$^{-2}$; see eg. \citet{ant93}], this would
effectively block out any soft X-ray or UV emission from the disk, yet
significant amounts of both are observed. Therefore if our modeling
is correct, either the molecular torus is geometrically-thin,
optically-thin, patchy or absent altogether, or the planes of the AD
and the torus are more orthogonal than coincident.

\citet{bol97} suggested that NLS1s in general may contain disks
viewed relatively {\sl face-on} which would contradict the results of
our modeling. However, they based this hypothesis on the high levels
of variability seen in many NLS1s (and in IRAS~13224--3809 in
particular) and \rej\ has shown only relatively small changes over
long time-scales. Thus while the variability may be related to the
inclination of the disk, our model fits suggest that the ultra-soft
X-ray emission, the low-velocity BLR and the direct or indirect link
between the two, may not be.

\subsection{The optical/UV `power-law' continuum}

When the FOS data are combined with the spectrum of the nuclear
optical continuum (see Section 3; Fig. 5), we find a very flat,
power-law-like component with a spectral index, $\alpha\sim$1,
covering the 8000~\AA\ to 1200~\AA\ range.  There is no sign of any
rise towards the blue in this region (which is characteristic of the
BBB), and the emission from the host galaxy of the AGN has been
removed. The residual component appears to be non-thermal in nature.
It cannot be due to the sum of thermal dust components because dust
grains sublime at 10$^5$~K and their emission is negligible at
wavelengths short-ward of 1$\mu$m. Therefore, this represents the first
unambiguous detection of a non-thermal optical/UV continuum component
in any AGN.

When modeling the optical to X-ray data, it was necessary to assume
an underlying optical to X-ray power-law, in addition to the accretion
disc component, to fit the optical/UV continuum {\sl and} the
hardening of the X-ray spectrum towards harder energies. The slope of
this power-law was $\alpha$=1.15, slightly softer than that in the
optical/UV alone. The resulting deficit in the UV between the data and
the power-law was well-modeled by the accretion disc component (see
Figs 7 and 8). Thus the data and the modeling presented here, are
consistent with the presence of a single power-law component which
underlies the optical to X-ray spectrum of \rej.

The existence of a single power-law component connecting the IR to the
X-rays has been suggested before \citep{mal84,elv86,car87}, although
the `bare' power-law, such as we have observed in \rej, has not (to
our knowledge) been observed, due to the dominance of the `thermal'
components in the optical and IR, respectively. The IR to UV continuum
is generally thought to be the sum of thermal components from dust and
the accretion disk. The hard X-ray power-law is considered to be
physically unrelated and produced by inverse Compton scattering of
blackbody photons from the disk in a surrounding hot corona. Previous
claims of a continuous IR-to-X-ray power-law component 
have been based on correlations between IR and X-ray
luminosities \citep{mal84,car87} and extrapolations of the medium
X-ray spectrum into the IR \citep{elv86}. In addition, \citet{fer00}
found evidence for a stable, underlying red continuum component in the
optical spectra of PG Quasars. The direct observation of the
optical/UV power-law component in \rej\ makes this a prime target for
the study of broadband, non-thermal emission in Seyfert galaxies.

Using {\sl ROSAT}-PSPC, optical, {\sl IUE} and archival {\sl IRAS}
data, \citet{puc95} also discussed the possibility of an
underlying IR to X-ray power-law component in \rej. These data
suggested a softer slope to the power-law however ($\alpha\sim$1.4),
due to the lack of any constraint in hard X-rays ($E>$2\keV) and
probable contamination of the optical spectrum by the host galaxy.

The similarity of this component to that observed in BL~Lac objects
led \citet{puc95} to hypothesize that there may be some kind of
`mini-BL~Lac' activity in \rej.  Although subsequent optical
spectropolarimetry of \rej\ revealed an upper limit on the
polarization of the continuum of 0.4\% \citep{bre98}, the
`duty cycle' of X-ray selected BL~Lac objects (\ie the fraction of
time spent in periods of significant variability and polarization) is
$\sim$40\% or less \citep{jan94,hei98}. Thus it is possible
that \rej\ does have a BL~Lac component which dominates in the
optical/UV, but is currently in a quiescent state.

%See also Urry and Padovani (1995), PASP, 107, 803

\section{Conclusions}

We have presented a comprehensive study of the narrow-line Seyfert 1
galaxy \rej, spanning the optical to X-rays and six years of
observations. Focusing on quasi-simultaneous \sax, \hst, \euve\ and
optical data taken in early 1997, we have {\sl (1)} measured the soft
and hard X-ray components simultaneously for the first time using
\sax; {\sl (2)} placed tight constraints on the form of 
the big blue bump using \hst, \euve\ and \sax; {\sl
(3)} separated out the nuclear component from the host galaxy in the
optical; and {\sl (4)} fitted the optical to X-ray nuclear continuum
with the combination of an accretion disk and an underlying power-law,
to constrain the black hole mass, accretion rate and inclination of
the accretion disk in \rej.

Coverage of the widest possible range in X-rays, particularly at soft
energies, is of prime importance for NLS1s, because of their strong
ultra-soft components. These \sax\ data, which cover the 0.1 to
10~\keV\ range, provide significant constraints on the shape, and thus
the physical nature, of the X-ray spectrum in \rej.  Our analysis of
the \sax\ spectrum reveals a significant hardening of the X-ray
spectrum above $\sim$3~\keV\ and that two blackbodies, with $kT_{\rm
eff}$=60~\eV\ and 160~\eV\ best represent the ultra-soft X-ray
component, consistent with multi-temperature emission from an
optically-thick accretion disc. We do not find any compelling evidence
to suggest that the hard X-ray power-law in \rej\ is significantly
softer than that of non-NLS1 AGN. There is the suggestion of a
low-energy flattening in the \sax\ data, which was also observed in
the earlier \rosat\ PSPC spectrum. This in turn implies a high
temperature for the inner edge of the accretion disk, and thus a low
black hole mass (which is also borne out by the modeling).

The results of fitting the optical to X-ray nuclear continuum have
shown that the data prefer relatively high mass accretion rates
($\sim$0.3 to 0.7 $\dot M_{\rm Edd}$) onto a low mass black hole
(M$_{bh}\sim$2-10$\times 10^6$ M$_{\odot}$) at high inclination angles (ie.
preferably edge-on, $\sim 60-70^\circ$ away from the axis of the
disc). The very low intrinsic columns implied by the X-ray fits
suggest that any molecular torus must either be optically thin,
geometrically thin, or lie out of the plane of the accretion disk so
that it does not obscure our line of sight. 

The optical to X-ray nuclear continuum {\sl requires} the presence of
a flat, power-law-like component in the optical/UV. This is consistent
with an extrapolation to the hard X-ray spectrum, although an
independent origin for the hard X-ray power-law cannot be ruled
out. The existence of such a component has been proposed before, but
we believe that this is the first direct observation, by virtue of the
unusually hot big blue bump in \rej. It appears to be stable and
unpolarized and we speculate that it may be a BL~Lac-type component
which is currently in the quiescent stage of its duty-cycle.

\acknowledgments

We thank the anonymous referee for comments which improved the paper.
This research has made use of data obtained through the High Energy
Astrophysics Science Archive Research Center Online Service, provided
by the NASA/Goddard Space Flight Center. This work is partly supported
by the Italian Space Agency, contract ARS/98--119 and by the Ministry
for University and Research (MURST) under grant COFIN--98--02--32. This
research has made use of data obtained from the Leicester Database and
Archive Service at the Department of Physics and Astronomy, Leicester
University, UK. The WHT is operated on the island of La Palma by the 
Isaac Newton Group in the Spanish Observatorio del Roque de los 
Muchachos of the Instituto de Astrofisica de Canarias.

%% Generally speaking, only the figure captions, and not the figures
%% themselves, are included in electronic manuscript submissions.
%% Use \figcaption to format your figure captions. They should begin on a
%% new page.

%\clearpage

%% No more than seven \figcaption commands are allowed per page,
%% so if you have more than seven captions, insert a \clearpage

%% There must be a \figcaption command for each legend. Key the text of the
%% legend and the optional \label in curly braces. If you wish, you may
%% include the name of the corresponding figure file in square brackets.
%% The label is for identification purposes only. It will not insert the
%% figures themselves into the document.
%% If you want to include your art in the paper, use \plotone.
%% Refer to the on-line documentation for details.

\begin{figure}
\epsscale{1}
\plotone{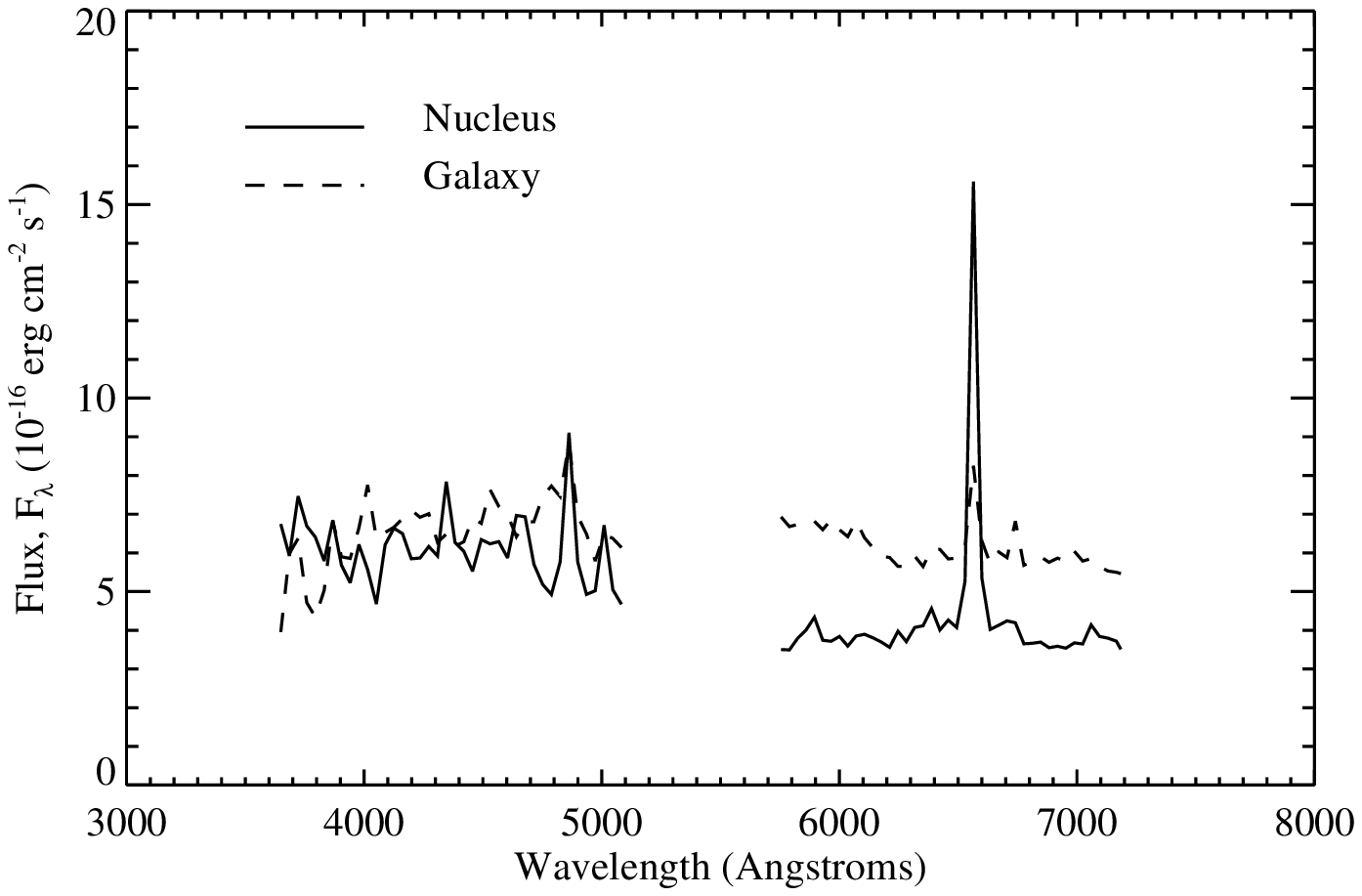}
\caption{The separate nuclear (plotted as a solid line) and galactic 
(dashed line) components of the optical ({\sl WHT}) spectrum of \rej.}
\end{figure}

%\clearpage

\begin{figure}
\epsscale{1}
\plotone{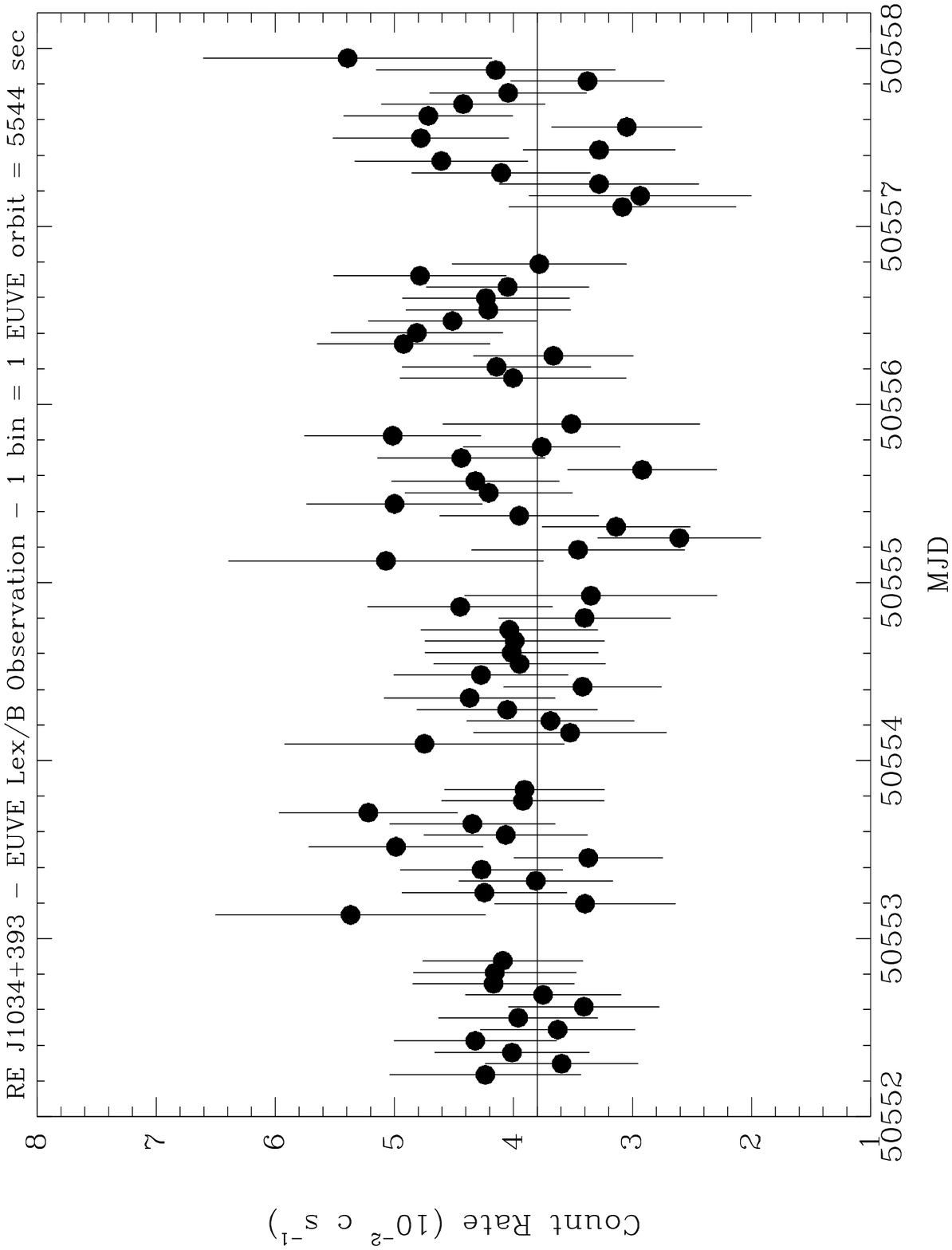}
\caption{The \euve\ light-curve of \rej.}
\end{figure}

%\clearpage

\begin{figure}
\epsscale{1}
\plotone{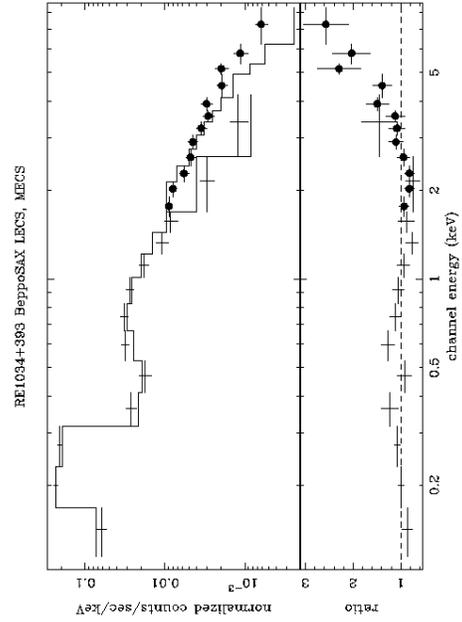}
\caption{The 0.1-10 keV BeppoSAX LECS, MECS spectra of \rej\
fitted with a power-law model reduced at low energy by Galactic
absorption. Lower panel shows the ratio between data and model. Error
bars are 1$\sigma$ (68\%).}
\end{figure}

%\clearpage

\begin{figure}
\epsscale{1}
\plotone{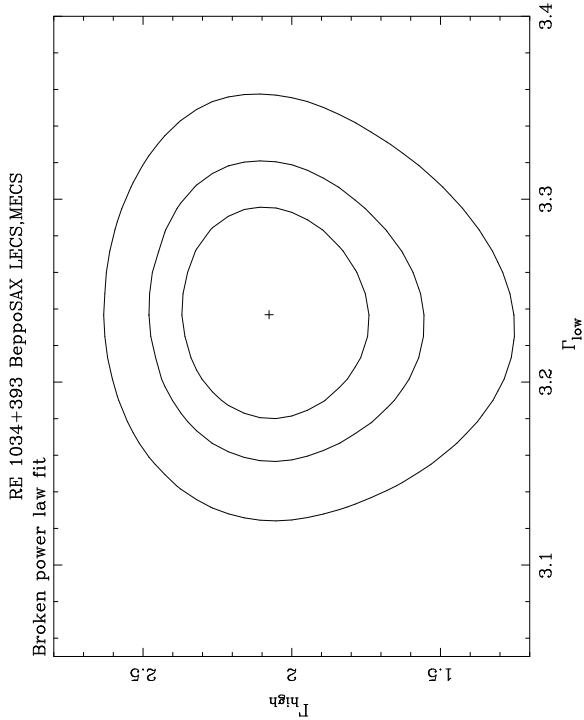}
\caption{The $\Gamma_{low}-\Gamma_{high}$ 
contour plot for a  broken power-law
fit to the LECS and MECS 0.1-10~\keV\ data of \rej. Contours are drawn
at 68\%, 90\% and 99\% significance for two interesting parameters.}
\end{figure}

%\clearpage

\begin{figure}
\epsscale{1}
%\plotone{opt_xray.ps}
\plotone{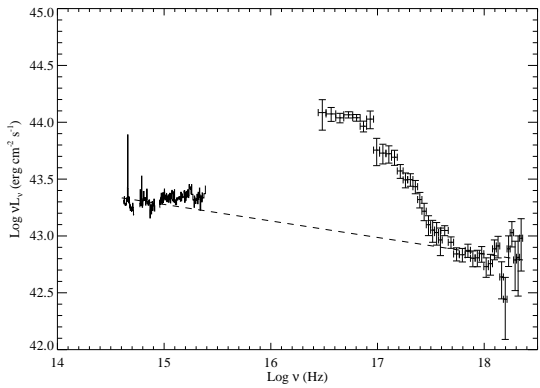}
\caption{The optical to X-ray spectrum of \rej, combining
the WHT, {\sl HST}-FOS and {\sl Beppo-SAX} data. The underlying
optical to X-ray power-law continuum used in the fits is also 
shown as a dashed line. Errors are 1$\sigma$ (68\%).}
\end{figure}

%\clearpage

\begin{figure}
\epsscale{1}
%\plotone{grids_jul00.ps}
\plotone{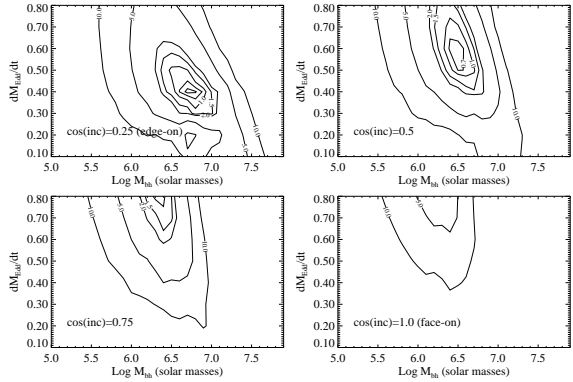}
\caption{$\chi_\nu^2$ contour plots for the AD plus power-law models 
when compared with the optical, UV and X-ray data for \rej.}
\end{figure}

%\clearpage

\begin{figure}
\epsscale{1}
\plotone{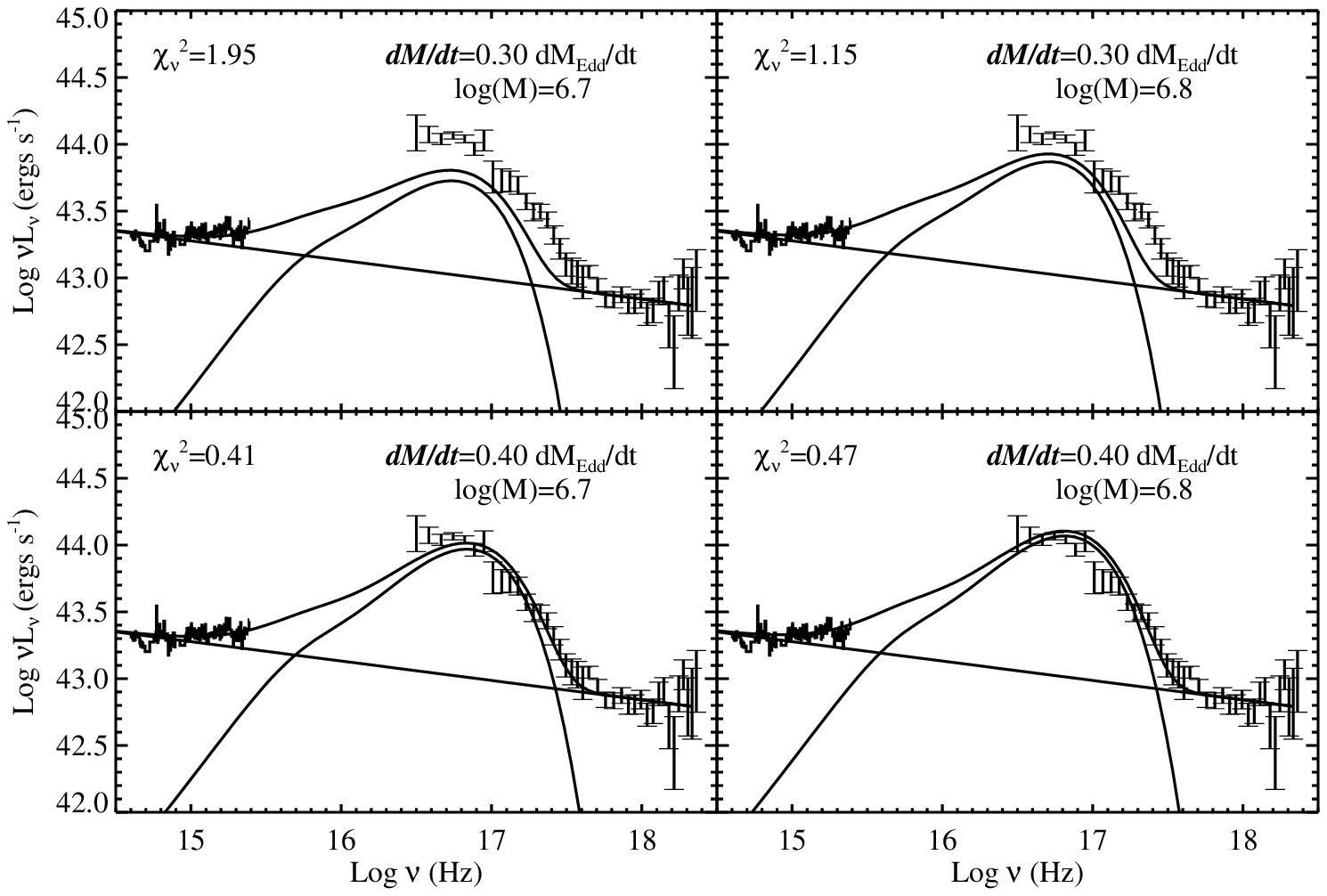}
\caption{AD plus power-law fits to the optical to X-ray 
data at an angle of inclination, cos(inc)=0.25.}
\end{figure}

%\clearpage

\begin{figure}
\epsscale{1}
\plotone{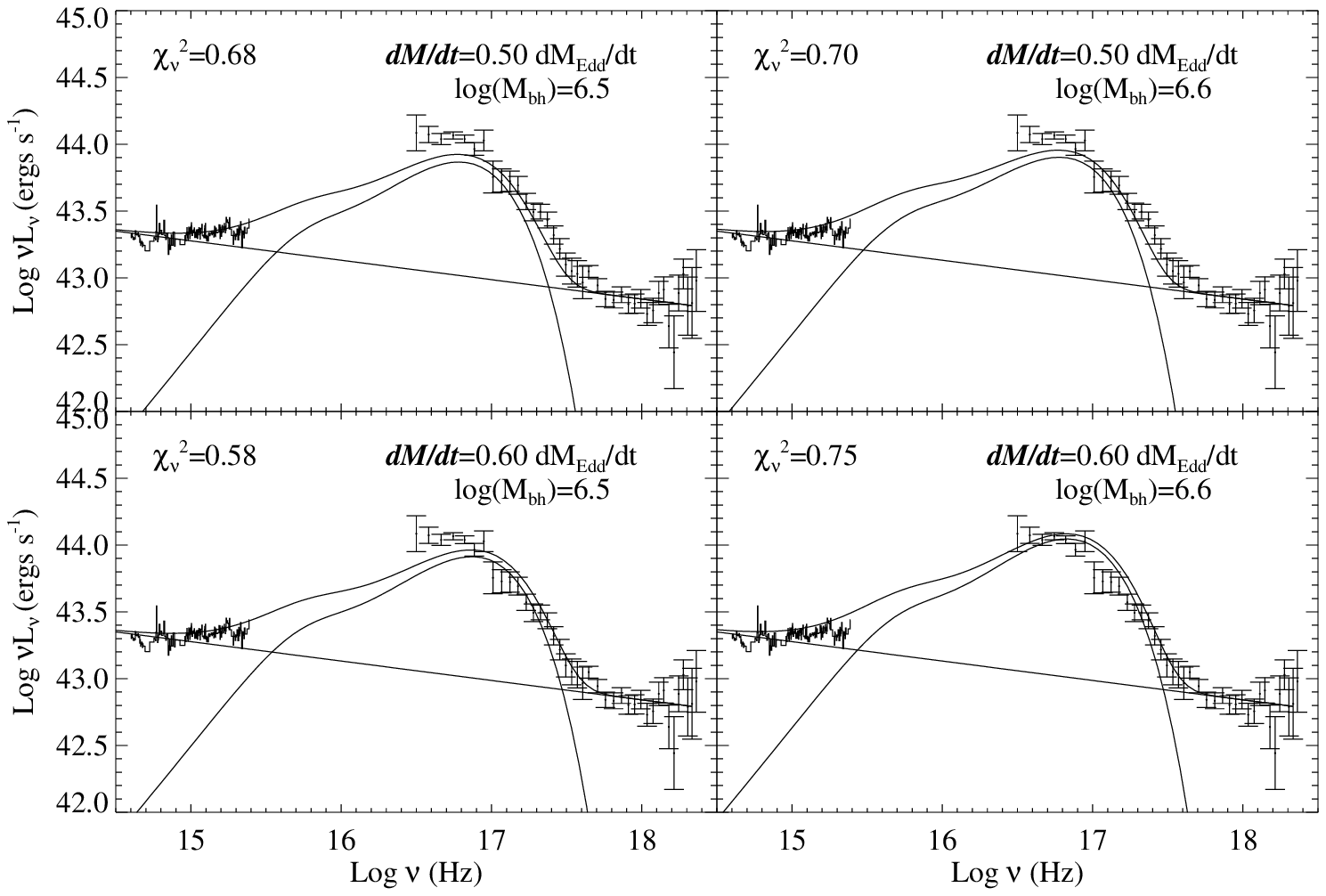}
\caption{AD plus power-law fits to the optical to X-ray 
data at an angle of inclination, cos(inc)=0.5.}
\end{figure}

\clearpage

\begin{deluxetable}{llcccc}
\tabletypesize{\scriptsize}
\tablewidth{0pt}
\tablecaption{Log of EUV and X-ray observations}
\tablehead{
\colhead{Date}           & 
\colhead{Instrument}      &
\colhead{Observation No.}          & 
\colhead{Exposure time (ks)}  &
\colhead{Count rate 10$^{-2}$ count s$^{-1}$}          & 
\colhead{Reference}}
\startdata
1991 Nov 18    & {\sl ROSAT} PSPC & {\sc RP}700551 & 330 & 3 & (1)\cr
1992-1993      & \multispan2{{\sl EUVE} All-Sky Survey}\hfil 
                                              & 1.6 & 2.3$\pm$0.7 & (2) \cr
1993 Dec 27    & {\sl EUVE} DS/S  &               & 5 & 2.9$\pm$0.3 & (3) \cr
1994 Nov 20    & {\sl ROSAT} HRI & {\sc RH}701982 & 1.6 & 49$\pm$2   & (4) \cr
1994 Nov 19   & {\sl ASCA} GIS & 72020000 & 26.1 & 3.0$\pm$0.1 (GIS2) & (4) \cr
1994 Nov 19   &                &          & 25.7 & 3.4$\pm$0.2 (GIS3) & (4) \cr
1994 Nov 19   & {\sl ASCA} SIS & 72020000 & 28.2 & 9.1$\pm$0.3 (SIS0) & (4) \cr
1994 Nov 19   &                &          & 28.2 & 7.6$\pm$0.2 (SIS1) & (4) \cr
1995 May 18   & {\sl ASCA} GIS & 72020010 & 10.5 & 2.5$\pm$0.2 (GIS2) & (4)\cr
              &                &          & 10.3 & 3.3$\pm$0.2 (GIS3) & (4)\cr
1995 May 18   & {\sl ASCA} SIS & 72020010 &  8.9 & 8.4$\pm$0.4 (SIS0) & (4)\cr
              &                &          &  9.0 & 6.5$\pm$0.3 (SIS1) & (4)\cr 
1996 Nov 12    & {\sl ROSAT} HRI & {\sc RP}703893 & 30 & 61$\pm$2 & (4) \cr 
1997 Apr 17-18 & {\sl SAX} LECS  & 50172004 & 21.7 & 6.9$\pm$0.2 & (4) \cr 
1997 Apr 17-18 & {\sl SAX} MECS  & 50172004 & 43.2 & 2.16$\pm$0.08 & (4) \cr
1997 Apr 14-20 & {\sl EUVE} DS/S & & 97.1 & 3.8$\pm$0.1 & (4) \cr 
\enddata
\tablenotetext{a}{(1) Puchnarewicz et al (1995); (2) Fruscione et al. (1996);
(3) Fruscione, private communication; (4) this paper; 
{\sl ASCA} GIS count rates are over 1.0-10~\keV, 
SIS count rates 0.7-10~\keV}
 
\end{deluxetable}

\begin{deluxetable}{lccccl}
\tabletypesize{\scriptsize}
\tablewidth{0pt}
\tablecaption{Fits to {\sl ASCA} spectra}
\tablehead{
\colhead{Model}           & 
\colhead{intrinsic $n_H$}      &
\colhead{$\alpha_S$ or kT (\eV)}          & 
\colhead{second BB kT (\eV)}  &
\colhead{$\alpha_H$}          & 
\colhead{$\chi^2$/dof}}
\startdata
\multispan3{\bf 72020000 Observation (1994 November)\hfil}\cr
\noalign{\smallskip}
One power-law &
0.0$<$0.51    &  2.04$\pm$0.09     & &                  & 519/410  \cr
       &      &  1.05$\times 10^{-3}$                               \cr
\noalign{\smallskip}
Two power-law &
0.0$<$28.6    &  3.43$_{-0.63}^{+2.08}$  & & 0.79$_{-0.62}^{+0.46}$ 
              & 363/408  \cr
       &      &  7.32$\times10^{-4}$  & & 3.07$\times 10^{-4}$          \cr
\noalign{\smallskip}
Broken power-law  &
0.0$<$7.4     &  2.61$_{-0.23}^{+0.45}$  & & 1.27$_{-0.28}^{+0.20}$ 
              & 375/408  \cr
       &      &  1.07$\times 10^{-3}$    & & break 1.51~\keV\             \cr
\noalign{\smallskip}
Brems+power-law   &
0.0$<$16.7    &  250$_{-66}^{+62}$ & & 1.18$\pm$0.25 & 367/408  \cr
       &      &  1.68$\times 10^{-2}$   & & 5.54$\times 10^{-4}$   \cr
\noalign{\smallskip}
BB + power-law &   
0.0$<$3.7     &  127$_{-19}^{+17}$ & & 1.29$_{-0.20}^{+0.19}$ &
371/408  \cr
       &      &  4.68$\times 10^{-5}$  & & 6.25$\times 10^{-4}$            \cr
\noalign{\smallskip}
2BB + power-law   & 
0.0$<$40      &  122$_{-62}^{+28}$ & 323$_{-183}^{+157}$  
              & 0.65$_{-1.25}^{+0.65}$
                                                        & 362/406  \cr
       &      &  6.43$\times 10^{-5}$ & 8.77$\times 10^{-6}$ 
              &  2.57$\times 10^{-4}$  \cr
\noalign{\bigskip}
\multispan3{\bf 72020010 Observation (1995 May)\hfil}\cr
\noalign{\smallskip}
Model                  &           
intrinsic $n_H$        &
$\alpha_S$ or kT (\eV) &
                       &
$\alpha_H$             &
$\chi^2$/dof       \cr
                       &           
                       &
normalization          &
normalization          &
normalization          &
                       \cr
\noalign{\smallskip}
One power-law    &
0.0$<$2.2       &  1.98$\pm$0.13     & &                  & 151/156  \cr
          &      &  1.04$\times 10^{-3}$                   \cr
\noalign{\smallskip}
Two power-law &
\multispan5{\hfil Too poorly constrained - no improvement over one
power-law
\hfil}\cr
\noalign{\smallskip}
Broken power-law  &
37$_{-35}^{+67}$ &  4.31$_{-1.83}^{+3.62}$  & & 1.79$_{-0.35}^{+0.62}$ 
                 & 120/154  \cr
       &         &  2.54$\times 10^{-3}$      & & break 1.40~\keV\  \cr
\noalign{\smallskip}
Brems+power-law   &  56$_{-51}^{+56}$    &  151$_{-48}^{+117}$     & 
                 & 1.82$_{-0.49}^{+0.60}$ & 123/154  \cr
                 &      &  1.16              & & 1.15$\times 10^{-3}$     \cr
\noalign{\smallskip}
BB + power-law &   
46$<$99       &  104$_{-23}^{+58}$  & & 1.79$_{-0.44}^{+0.55}$ & 124/154  \cr
       &      &  6.05$\times 10^{-4}$ & & 1.11$\times 10^{-3}$             \cr
\noalign{\smallskip}
2BB + power-law   & \multispan5{\hfil Too poorly constrained - no improvement over one blackbody\hfil}\cr
\enddata
\tablenotetext{a}{Intrinsic n$_H$ is the column density local to the AGN,
{\sl in addition} to a Galactic absorbing column density fixed at
1.5$\times 10^{20}$ cm$^{-2}$, and corrected for redshift.  Limits
(given in brackets) are 90\%. Power-law indices, $\alpha$, are
defined such that $F_\nu\propto\nu^{-\alpha}$. Bremsstrahlung and
blackbody temperatures are given in eV and have been corrected for
redshift. Power-law and broken power-law normalizations are in units
of photons \keV$^{-1}$ cm$^{-2}$ s$^{-1}$ at 1~\keV; bremsstrahlung
normalization = ${3.02\times 10^{-15}}/(4\pi D^2)\int n_e n_I
dV$, where $n_e$ is the electron density in cm$^{-3}$, $n_I$ is the
ion density in cm$^{-3}$ and D is the distance to the source in cm;
blackbody normalization = $L_{\rm 39}/{D_{\rm 10}^2}$, where $L_{\rm
39}$ is the source luminosity in units of 10$^{39}$ erg s$^{-1}$ and
$D_{\rm 10}$ is the distance to the source in units of 10 kpc.}
\end{deluxetable}

%\clearpage

\begin{deluxetable}{lllll}
\tablewidth{0pt}
\tablecaption{Log of Beppo-SAX observations}
\tablehead{
\colhead{Dates}           & 
\colhead{Exposure}      &
\colhead{ }      &
\colhead{Rate (10$^{-2}$ c/s)} &
\colhead{ }      }
\startdata
               & LECS         &  MECS    &  LECS         &  MECS \cr
18-19 Apr 1997 & 21.7ks       &  43.2ks  &  6.9$\pm$0.2  &  2.16$\pm$0.08 \cr
\enddata
\end{deluxetable}

%\clearpage

\begin{deluxetable}{lccccc}
\tablewidth{0pt}
\tablecaption{Fits to Beppo-SAX data}
\tablehead{
\colhead{Model}           & 
\colhead{$N_H^a$}      &
\colhead{$\alpha_S$ or kT$^b$} &
\colhead{$E_{break}^c$ or kT$^b$}      &
\colhead{$\alpha_H$}      &
\colhead{$\chi^2$(dof)}}       
\startdata
\multispan{6}{\bf LECS plus MECS  0.1-10~\keV \hfil} \cr
power-law   & 1.75$\pm$0.35          & 2.27$\pm$0.14   & ...             
            & ...                    & 120.2/92 \cr
broken p.l. & 2.4$\pm$0.5            & 2.54$\pm$0.16   & $E_b=2.3\pm$0.5 
            & 1.13$^{+0.29}_{-0.35}$ & 70.0/90  \cr
broken p.l. & 1.5(F) & 2.24$\pm$0.07 & $E_b=2.5\pm$0.6 & 1.08$\pm$0.32 
            & 83.2/91 \cr
bb$+$p.l.   & 0.5$^{+0.5}_{-0.35}$   & kT=100$\pm$10   & ... 
            & 1.50$\pm$0.26          & 95.1/90  \cr
bb$+$p.l.   & 1.5(F)                 & kT=82$^{+13}_{-8}$ & ... 
            & 1.84$\pm$0.25          & 104.6/91 \cr 
bb$+$bb$+$p.l. & 1.5(F)              & kT$_1$=55$\pm$10 
            & kT$_2$=155$^{+26}_{-20}$ & 1.19$\pm$0.24 &  65.7/89 \cr
\noalign{\smallskip}\cr
\multispan{6}{\bf MECS 2-10 keV\hfil} \cr
power-law   & 1.5(F)                 & 1.42$\pm0.26$   & ... 
            & ...                    & 31.5/45 \cr
\noalign{\smallskip}\cr
\multispan{6}{\bf LECS 0.1-1.5 keV\hfil} \cr
power-law   & 2.3$\pm$0.5            & 2.45$\pm$0.20   & ... 
            & ...                    & 29.9/41 \cr
\enddata
\tablenotetext{1}{$^a$ Absorbing column density in units of $10^{20}$
cm$^{-2}$; 
$^b$ Temperatures in eV. 
$^c$ Break energies in \keV; (F) frozen parameter. Errors are 90\%.}
\end{deluxetable}

%\clearpage

\end{document}